\documentclass[twoside]{article}
\usepackage{qic}
\usepackage{lineno,hyperref,epstopdf}
\usepackage{graphics}

\begin{document}
\setlength{\textheight}{8.0truein}    %FOR 2ND PAGE ONWARDS

\runninghead{Title  $\ldots$}
            {Author(s) $\ldots$}

\normalsize\textlineskip
\thispagestyle{empty}
\setcounter{page}{1}

%\copyrightheading{Vol.}{No.}{Year}{Page Nos.}
\copyrightheading{0}{0}{2003}{000--000}

\vspace*{0.88truein}

\alphfootnote

\fpage{1}

\centerline{\bf Learning quantum annealing}
\vspace*{0.37truein}
\centerline{\footnotesize E.C. Behrman}
\vspace*{0.015truein}
\centerline{\footnotesize\it Department of Mathematics and Physics, Wichita State University}
\baselineskip=10pt
\centerline{\footnotesize\it Wichita, KS 67260-0033,USA}
\vspace*{10pt}
\vspace*{10pt}
\centerline{\footnotesize J.E. Steck and M.A. Moustafa}
\vspace*{0.015truein}
\centerline{\footnotesize\it Department of Aerospace Engineering, Wichita State University}
\baselineskip=10pt
\centerline{\footnotesize\it Wichita, KS 67260-0044, USA}

\vspace*{0.21truein}

\abstracts{We propose and develop a new procedure, whereby a quantum system can learn to anneal to a desired ground state. We demonstrate successful learning to produce an entangled state for a two-qubit system, then demonstrate generalizability to larger systems.   The amount of additional learning necessary decreases as the size of the system increases. Because current technologies limit measurement of the states of quantum annealing machines to determination of the average spin at each site, we then construct a ``broken pathway'' between the initial and desired states, at each step of which the average spins are nonzero, and show successful learning of that pathway. Using this technique we show we can direct annealing to multiqubit GHZ and W states, and verify that we have done so. Because quantum neural networks are robust to noise and decoherence we expect our method to be readily implemented experimentally; we show some preliminary results which support this.
}{}{}

\vspace*{10pt}

\vspace*{3pt}
\communicate{to be filled by the Editorial}

\noindent
\vspace*{1pt}\textlineskip	%) USE THIS MEASUREMENT WHEN THERE IS
\section{Introduction\label{intro}}        %) A SECTION HEADING
\vspace*{-0.5pt}
\noindent
Most efforts to design a quantum computer over the past several decades have used a circuit model. In this model, a problem is solved algorithmically, using circuits built up of simple blocks. A minimum operation set of these blocks was established early\cite{needent}, and many physical implementations were then proposed as candidates for the universal quantum computer: any simple, manipulable quantum system that could be made to perform that minimum set. But the leap to a macroscopic computer which could solve interesting problems proved difficult. Solid state approaches using quantum dots and/or impurities \cite{solidstate} hold out the promise of utilizing a great deal of the technology developed over the last century, yet problems remain. Optical computing\cite{optical} is another approach, but has problems with coupling efficiency and fault tolerance. 

A different model is a kind of analog computer called a quantum annealing machine\cite{qannealing}, that solves binary optimization problems.  Systems are now being built\cite{dwave} with hundreds of qubits. The basic idea is as follows: one maps one's problem onto that of finding the energy minimum of an Ising model, the  Hamiltonian of this model is found, and then the machine is initialized to the ``flat'' (equal amplitudes of all basis states) state and slowly annealed to the problem Hamiltonian. If all goes well, the state to which the machine has relaxed, using tunneling, is now the solution to the problem. With current technology, that complete final quantum state cannot be read out, but the average spin of each qubit can be, and sometimes this is sufficient. 

It is not at all clear that these systems use the full power of quantum computing, for which entangled states are necessary\cite{needent,lanting}. Multiqubit entanglement is common in nature but, depending on the physical setup, it is not always easy to produce it in a controlled manner\cite{kimble, entlight, huang} or to maintain it\cite{huang,novotny,hou,hu,xiang-ping}. In previous work, we have developed an entanglement indicator\cite{behrmanqic}, corrected it for anomalous oscillation\cite{behrmanieee}, and extended it to multiqubit systems\cite{behrmannabic,behrmanmulti}. As the size of the system grows the amount of additional training necessary diminishes, a kind of ``bootstrapping'' effect\cite{efron,wasserman}. Thus, unlike other methods which require knowledge or reconstruction of the density matrix \cite{bennett2, wootters, ghz}, our learning and bootstrapping methods potentially may be of general applicability even to large-scale quantum computers, once they are built. We have also shown\cite{behrmannoise} that the indicator is robust to noise and decoherence. For both these reasons the quantum learning approach has advantages in scaleup.

Clearly it would be of great interest to have a systematic way of creating entanglement in a many-qubit system. Previous authors have explored doing this with Bose-Einstein condensates\cite{cirac}. Here, we propose extending the capability of solid state annealing machines, by showing that specific, desired, states can be prepared by the annealing process. This would mean that a quantum annealing array could be used as an algorithmic computer, or, more generally, as a quantum neural computer or quantum neural network (QNN). We call this ``learning quantum annealing.'' That is, instead of specifying the Hamiltonian and using the array to find the minimum energy, we do the inverse problem: We specify the desired state and find the Hamiltonian that will produce that state as its final state along the annealing pathway. We show, specifically, that  two-, three-, four-, five-, and six-qubit arrays, initialized to the flat state, can be made to anneal to the corresponding GHZ state, via a time sequence of coupling functions, which are found by using machine learning\cite {lecun, werbos}. Currently, the only experimentally measureable quantity is the average spin at any given site. Because the initial (flat) and the final (GHZ) states are both symmetric (i.e., the average spins are all zero), we need, in addition, some method for measuring whether we have in fact succeeded. We therefore show that our quantum learning technique can also be used along a pathway to the final state for which the average spin is not zero. Breaking that path into steps, then annealing to those intermediate steps followed by measurement of the spins at each step, will provide strong experimental evidence of the success of our technique.

To show that the technique is not limited to creating GHZ states, we also show that annealing paths to multiqubit pairwise entangled W states can be learned. In addition, we provide evidence that the method is robust to noise and to decoherence.

\vspace*{1pt}\textlineskip	%) USE THIS MEASUREMENT WHEN THERE IS
\section{Machine learning of annealing\label{qnn}}	        %) A SECTION HEADING
\vspace*{-0.5pt}
\noindent
We begin with the Schr\"{o}dinger equation:
\begin{equation}
\frac{d \rho}{dt} = \frac{1}{i \hbar}[H, \rho] 
\label{schr}
\end{equation}
where $\rho$ is the density matrix and $H$ is the Hamiltonian. We consider an N-qubit system whose Hamiltonian is
\begin{equation}
H   = \sum_{\alpha=1}^{N}  K_{\alpha} \sigma_{x\alpha} + \varepsilon_{\alpha} \sigma_{z\alpha} + \sum_{\alpha\neq\beta=1}^{N}\zeta_{\alpha\beta} \sigma_{z\alpha} \sigma_{z\beta} ,
\label{ham}
\end{equation}
where $\{ \sigma \}$ are the Pauli operators corresponding to each of the qubits, $\{K \}$ are the tunneling amplitudes, $\{ \varepsilon \}$  are the biases, and $\{ \zeta \}$, the qubit-qubit couplings. We choose the usual ``charge basis'', in which each qubit's state is given as up or down, +1 or -1, denoted by $|0\rangle$ and $|1\rangle$, respectively.  For a system of N qubits there are thus $2^{N}$ states, each labelled by a bit string each of whose numbers corresponds to the state of each qubit, in order. The amplitude for each qubit to tunnel to its opposing state (i.e., switch between the $|0\rangle$ and $|1\rangle$ states) is its $K$ value; each qubit has an external bias represented by its $\varepsilon$ value; and each qubit is coupled to each of the other qubits, with a strength represented by the appropriate $\zeta$ value. Note that, for example, the operator $\sigma_{xA} = \sigma_{x}\otimes I \: ... \: \otimes I$, where there are (N-1) outer products, acts nontrivially only on qubit A. 

The parameters $\{ K,\varepsilon,\zeta \}$ are allowed to vary with time and direct the time evolution of the system in the sense that, if one or more of them is changed, the way a given state will evolve in time will also change, because of Eqs.~(1)-(2). We use a quantum machine learning paradigm using quantum  backpropagation\cite{lecun} in time\cite{werbos} to find these parameter functions that produce desired quantum states. In previous work\cite{behrmanqic}, via machine learning, we successfully mapped an entanglement witness of the system's initial state, to a measurement at a final time $t_{f}$. Here, we wish instead to direct the time evolution while at the same time performing quantum annealing by lowering the temperature and/or reducing the tunneling amplitudes. Because the temperature is a measure of the energy available to the system, with appropriately learned parameters the system will anneal to the desired state; or, if the tunneling is reduced to zero, the system will be ``frozen'' into the state to which it has evolved. This is a kind of quantum control\cite{control}. 

Formally, the solution to Eq.(1) is given, for constant $H$ (we will allow $H$ to vary with time later), by 
\begin{equation}
\rho(t) = \exp (iHt/\hbar) \rho (0) \exp(-iHt/\hbar).
\label{schrsoln}
\end{equation}
We analytically continue the Schr\"{o}dinger equation to imaginary time, $t \rightarrow i\beta\hbar$, to find the density matrix as a function of temperature:
\begin{equation}
\rho(\beta) = \exp (-\beta H) \rho (0) \exp(\beta H),
\label{schrimag}
\end{equation}
where $\beta$ is the inverse temperature in units of Boltzmann's constant. We now split the dependence using the interaction representation: thinking of the two parts of the time evolution as being due to two parts of a joint Hamiltonian, one in real and one in imaginary time. Thus we integrate the Schr\"{o}dinger equation numerically in real time to find the time evolution of $\rho_{S}(t)$, the solution to Eq.(3), and find the temperature dependence with 
\begin{equation}
\rho_{I}(t, \beta) = \exp (-\beta H) \rho_{S}(t) \exp(\beta H).
\end{equation}

The above is straightforward. To implement our machine learning technique, we define a Lagrangian $L$ to be minimized as:

\begin{equation}
L = \frac{1}{2} |\rho_{I \texttt{des}} - \rho_{I}(t_{f})|^{2} + \int_{0}^{t_{f}} \lambda^{\dagger}(t) \exp(-\beta H) \left(\frac{\partial \rho_{S}}{\partial t} - \frac{i}{\hbar}[\rho_{S},H]\right)\exp(\beta H) \gamma(t)\, dt
\label{eqLagr}
\end{equation}
where the Lagrange multiplier vectors are $\lambda^{\dagger}(t)$ and $\gamma (t)$  (row and column, respectively),  
and $\rho_{I \texttt{des}}$ is the density matrix for the desired final state.  We now allow the ``weight'' parameters $\{ K,\varepsilon,\zeta \}$ to vary with time; this will change Eqs. (3)-(5) by the insertion of time-ordered integrals. In addition, we allow the inverse temperature $\beta$ also to vary with time, in order to make the time evolution process an annealing one.

We take the first variation of $L$ with respect to $\rho$, set it equal to zero, then integrate by parts to give the following equation which can be used to calculate the vector elements of the Lagrange multipliers (co-states) that will be used in the learning rule:

\begin{equation}
\gamma_{i} \frac{\partial \gamma_{j}}{\partial t} + \frac{\partial \lambda_{i}}{\partial t} \gamma_{j} - \frac{i}{\hbar} \sum_{k} \lambda_{k} H_{ki} \gamma_{j} + \frac{i}{\hbar} \sum_{k} \lambda_{i} H_{jk} \gamma_{k} = 0
\label{eqfive}
\end{equation}
with the boundary conditions at the final time $t_{f}$  given by
\noindent
\begin{equation}
-[\rho_{I \texttt{des}} - \rho_{I}(t_{f})]_{ji} + \lambda_{i}(t_{f})\gamma_{j}(t_{f}) = 0
\label{eqsix}
\end{equation}

The gradient descent learning rule is given by 
\noindent
\begin{equation}
w_{\texttt{new}} = w_{\texttt{old}} - \eta \frac{\partial L}{\partial w}
\label{eqseven}
\end{equation}
for each weight parameter $w$, where $\eta$ is the learning rate and
\noindent
\begin{eqnarray}
\frac{\partial L}{\partial w}  = \frac{\partial }{\partial w} \left\lbrace \int_{0}^{t_{f}} \lambda^{\dagger}(t) \left\lbrace \frac{\partial \rho_{I}}{\partial t} - \frac{i}{\hbar} [ \rho_{I},H] \right\rbrace \gamma(t) dt  \right\rbrace \\ \nonumber
 = \int_{0}^{t_{f}} \lambda^{\dagger}(t) \left\lbrace \frac{\partial}{\partial t}\beta(t)\left[ \rho_{I},\frac{\partial H}{\partial w}\right] - \frac{i}{\hbar}  \left[ \rho_{I},\frac{\partial H}{\partial w}\right] - \frac{i}{\hbar}\beta(t) \left[ \left[ \rho_{I}, \frac{\partial H}{\partial w}\right] , H \right] \right\rbrace \gamma(t) dt
\label{eqeight}
\end{eqnarray}

Here, $\beta(t)$, the inverse temperature, is a function of time: going from zero (at time zero) to the desired (high) annealed inverse temperature (at $t_{f}$.) For simplicity we take the dependence to be linear. 
Because of the Hermiticity of the Hamiltonian, $H$, and of the density matrix 
$\rho$, $\lambda_{i} \gamma_{j} = \lambda_{j} \gamma_{i} $ and the derivative of the 
Lagrangian, $L$, with respect to the weight, $w$, as given by Eq. (10), will be a real number.  
In addition, the two ``parts'' of the Hamiltonian commute (as is not usually the case with the interaction representation!) The derivative of the Lagrangian can also be written in terms of our earlier result\cite{behrmanqic} for zero $\beta$  as
\noindent
\begin{eqnarray}
\frac{\partial L}{\partial w} = \frac{\partial L}{\partial w}\vline_{\beta =0} + \int_{0}^{t_{f}} \beta(t) \lambda^{\dagger}(t)\left\lbrace \frac{\partial }{\partial t} \left[ \rho_{I} , \frac{\partial H}{\partial w}\right] - \frac{i}{\hbar} \left[  \left[ \rho_{I},\frac{\partial H}{\partial w}\right] , H \right] \right\rbrace \gamma(t)\, dt \\ \nonumber 
+ \frac{\beta_{f}}{t_{f}}\int_{0}^{t_{f}} \lambda^{\dagger}(t)\left[ \rho_{I} , \frac{\partial H}{\partial w}\right] \gamma(t)\, dt.
\label{eqeleven}
\end{eqnarray}
Note that the first correction term is of the same form as the original, but with the commutator playing the role of the density matrix.

Nonzero temperature is necessary in order that an entire continuum of equilibrium ground states be possible, but our technique is by no means limited to any particular annealing pathway. One advantage of simulations is that they include the probability amplitudes for kinetic as well as thermodynamic nearby states, which are experimentally accessible in (customary) multiple runs. 

Our method could also be called quantum system design through learning\cite{behrmanqic}, as machine learning is used to design an experimental quantum system to achieve a desired operational result, or quantum programming, as it is a method for choosing system parameters (software) to yield a desired result on a quantum computer.

\vspace*{1pt}\textlineskip	%) USE THIS MEASUREMENT WHEN THERE IS
\section{ Annealing to GHZ states} %) A SECTION HEADING
\vspace*{-0.5pt}
\noindent

The large SQuID arrays are normally initialized to what we call the ``flat'' state: a coherent equal superposition of all basis states. In the so-called ``charge'' basis, this looks like $\rho_{\texttt{flat}}= \frac{1}{2^{N}}\prod_{i=1}^{N} [|0\rangle + |1\rangle]_{i}\otimes [\langle 0| + \langle 1|]_{i}$. (For the simplest nontrivial case of two qubits, this is just a $4 \times 4$ matrix of ones.) To perform large scale calculations, {\it e.g.}, using Shor's error correction, it can be necessary to initialize the system in a fully entangled state. But how to do this? The (kinetic) transformation for a two qubit state is easy - indeed a textbook exercise - but for larger systems it would be of great value to be able to automate the process, to have the system itself learn to initialize in a fully entangled state.

We begin with the two qubit case. For this relatively simple system we trained only the coupling parameter function $\zeta$, setting the bias functions $\varepsilon_{A}(t)=\varepsilon_{B}(t)$ to zero for the entire time. In imitation of the annealing procedure used in quantum annealing\cite{qannealing,dwave}, we set the tunneling parameter  $K_{A}(t)=K_{B}(t)$ to a linearly decreasing function, from a large initial value to zero, during the first part of the time evolution; see Figure 1; the same tunneling parameter function will be used in all the calculations in this paper, for all qubits (with the exception of the preliminary single-function calculations shown in Figure 16.) Training details are summarized in Table 1. Figure 2 shows the error as a function of epoch (pass through the ``training set'', here a single ``training pair'' of the state of the system at the final time and temperature $\rho_{I}(t_f)$, and the desired (fully entangled) final state $\rho_{I \texttt{des}}$.) The system is initalized to the flat state, then evolved in time by integrating the Schr\"{o}dinger equation (Eq. (1)), using the parameter functions $\{ K_{A}(t)=K_{B}(t),\varepsilon_{A}(t)=\varepsilon_{B}(t)=0 \}$. $\zeta$ was initialized to be zero at all times. We compute $\rho_{I}$ at each timestep, using $\rho_{S}$ from the solution to Eq. (1), $\rho_{I}$ from Eq. (5), and the derivatives of the Lagrangian with respect to $\zeta$, using Eq. (10). We compute the error, then make small adjustments to $\zeta$ so as to decrease the total error; repeating the entire process multiple times until the error is small. As Figure 2 shows, the rms error approaches an asymptote very rapidly, and is equal to 0.00158 at only 50 epochs. The trained function $\zeta(t)$ is also shown, as is the annealing of the density matrix to the desired state as a function of time.

\begin{figure} [htbp]
\centerline{\epsfig{file=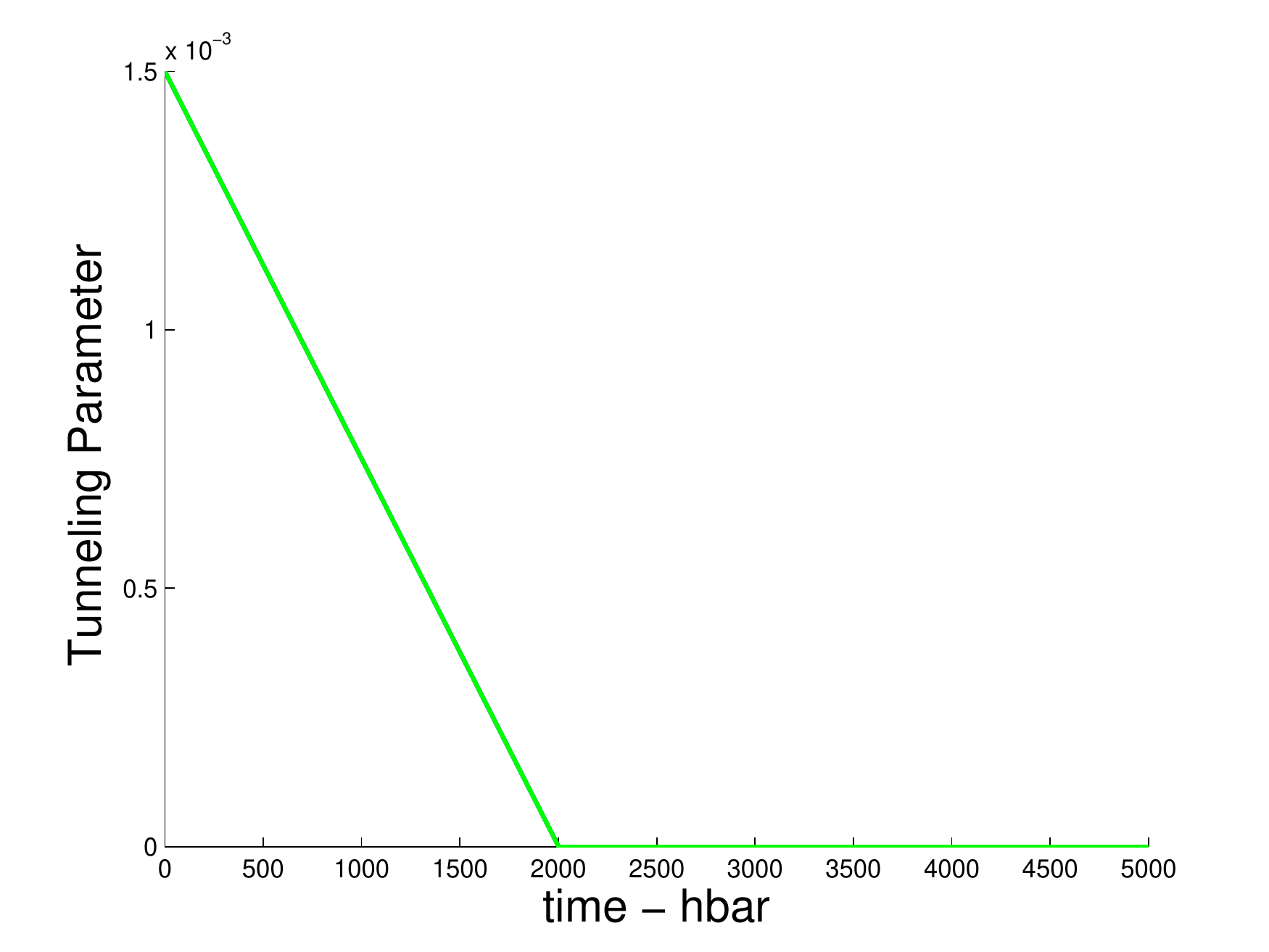, width=8.2cm}} %100 percent
\vspace*{13pt}
\fcaption{The tunneling parameter function $K(t) = K_{A}(t)=K_{B}(t)$, in GHz, as a function of time in units of  nsec per $\hbar$ . This function was set (not trained) to decrease (linearly) to zero, so as to keep the system in the desired state once it has annealed there. This is the usual procedure for commercial quantum annealing machines. This is the tunneling function we use for all the simulations in this paper (except Figure 16.) \label{K2}}
\end{figure}

\begin{table}
\tcaption{ Training data for flat to Bell. }
\centerline{\footnotesize\smalllineskip
\begin{tabular}{c c c }\\
\hline
\hline
Parameter (MHz) & Initial &  \\
\hline
$K_{A}$  & $1.5 \times 10^{-3}$ &  \\
$K_{B}$  & $1.5 \times 10^{-3}$ &   \\
$\zeta_{AB}$  & 0.0 &  \\
$\varepsilon_{A}$  & 0.0 &   \\
$\varepsilon_{B}$  & 0.0 &   \\
 \\
\hline \hline \\
$\beta_{f} = 2500; t_{f} = 5000\hbar; \Delta t = 5\hbar/2; \eta= 1.25 \times 10^{-05}$
\end{tabular}}
\end{table}

\begin{figure} [htbp]
%\centerline{\epsfig{file=2-eps-converted-to.pdf, width=8.2cm}} %100 percent
\centerline{\epsfig{file=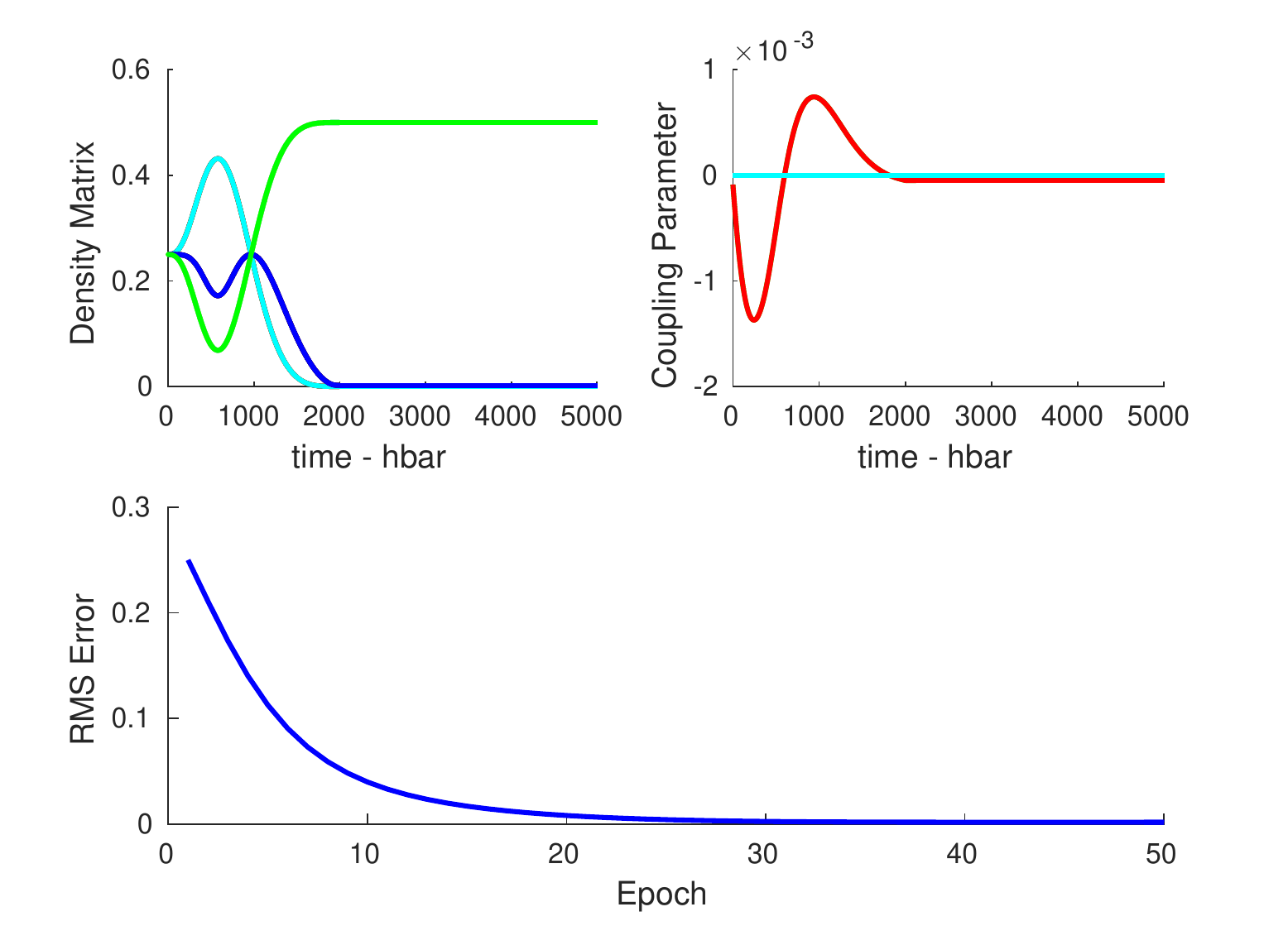, height=2.7in}} %100 percent
\vspace*{13pt}
\fcaption{Training the annealing of the two-qubit system from the flat state to the (fully entangled) Bell state. The top left graph shows the time evolution of (the absolute magnitude of the elements of) the density matrix $\rho_{I}(t)$, as a function of (annealing) time Note that, by symmetry, there are only three dissimilar numbers of the sixteen elements. Top right shows the trained parameter function $\zeta$, which directs the annealing, as a function of time. the bottom graph shows the root mean squared error for training as a function of epoch (pass through the training set). The asymptotic error was 0.00158.  The learning rate was $1.25 \times 10^{-05}$. }
\end{figure}

For the three-qubit system, we ``bootstrap'': We start from the trained $\zeta(t)$ function that we found for the two-qubit system, using it for the pairwise coupling function between the qubits, for each of the three pairs $AB$, $AC$, and $BC$. We keep the same bias function $\varepsilon=0$ and tunneling functions $K$ for each of the three qubits, and train from the flat initial state to the new desired state, the GHZ state $\frac{1}{\sqrt{2}}(|000\rangle+|111\rangle)$, using the same training rate as before. The new error as a function of epoch is shown in the bottom of Figure 3, which shows also the new trained coupling parameter function  $\zeta(t)$, and the time evolution of the density matrix to the desired state. Note the change in scale from the two-qubit training in Figure 2: the error starts out quite small, because most of the training has already taken place with the two-qubit system.

\begin{figure} [htbp]
\centerline{\epsfig{file=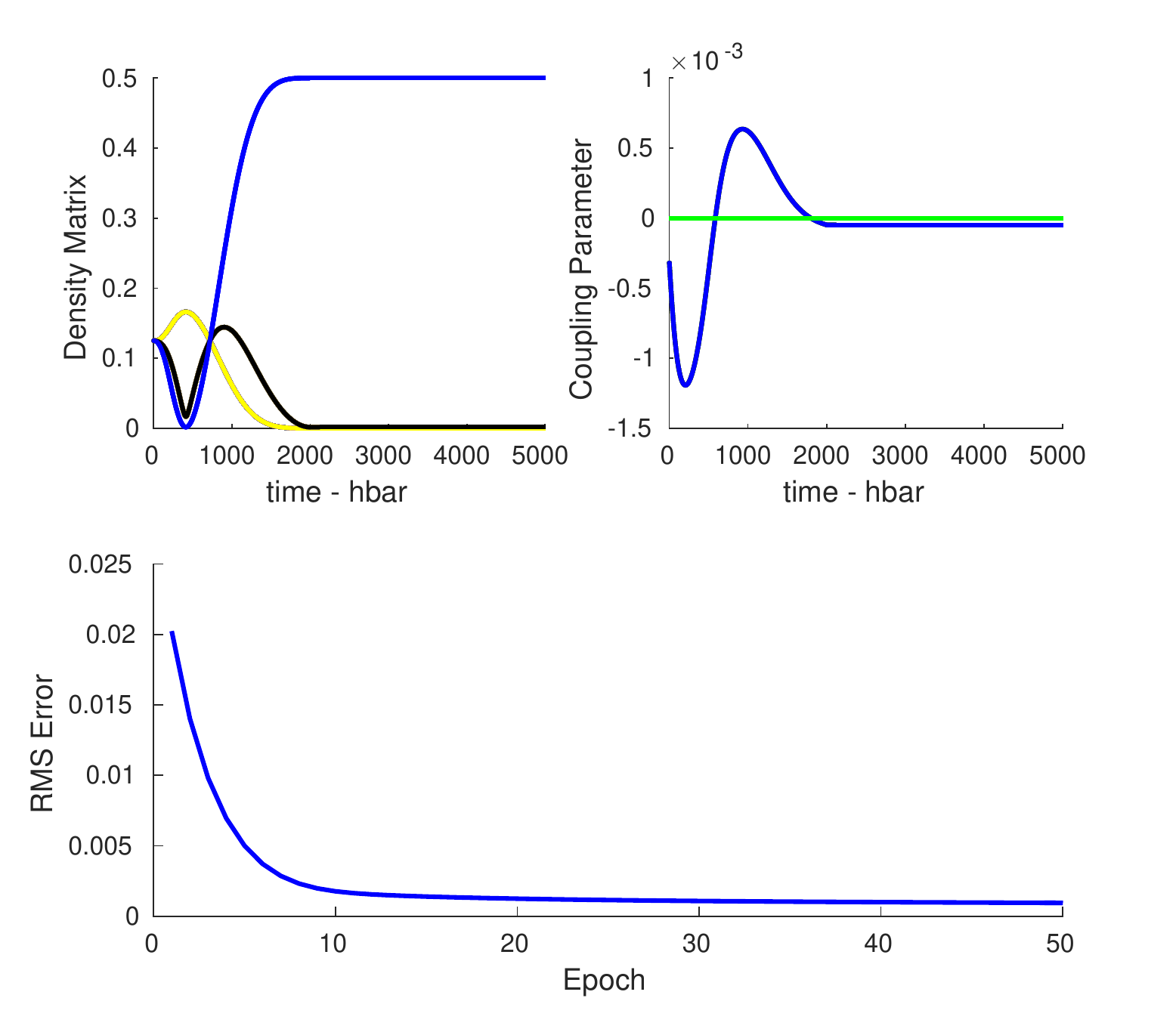, height=2.7in}} %100 percent
\vspace*{13pt}
\fcaption{Training the annealing of the three-qubit system from the flat state to the (fully entangled) GHZ state, starting from the trained two-qubit parameters. The top left graph shows the time evolution of (the absolute magnitude of the elements of) the density matrix $\rho_{I}(t)$, as a function of (annealing) time. Top right shows the trained parameter function $\zeta$, which directs the annealing, as a function of time, in units of GHz. The bottom graph shows the root mean squared error for training, as a function of epoch. The asymptotic error was 0.000944.  The learning rate was $1.25 \times 10^{-05}$. }
\end{figure}

We now successively boot to four-, five-, and six-qubits, each time starting from  the previously trained coupling function, and using the same training rate. Training for each is shown in Figures 4, 5, and 6. Note the change of scale each time as there is progressively less and less to learn. In all cases, except for booting from the three-qubit to the four, training takes place in fewer epochs. 

\begin{figure} [htbp]
\centerline{\epsfig{file=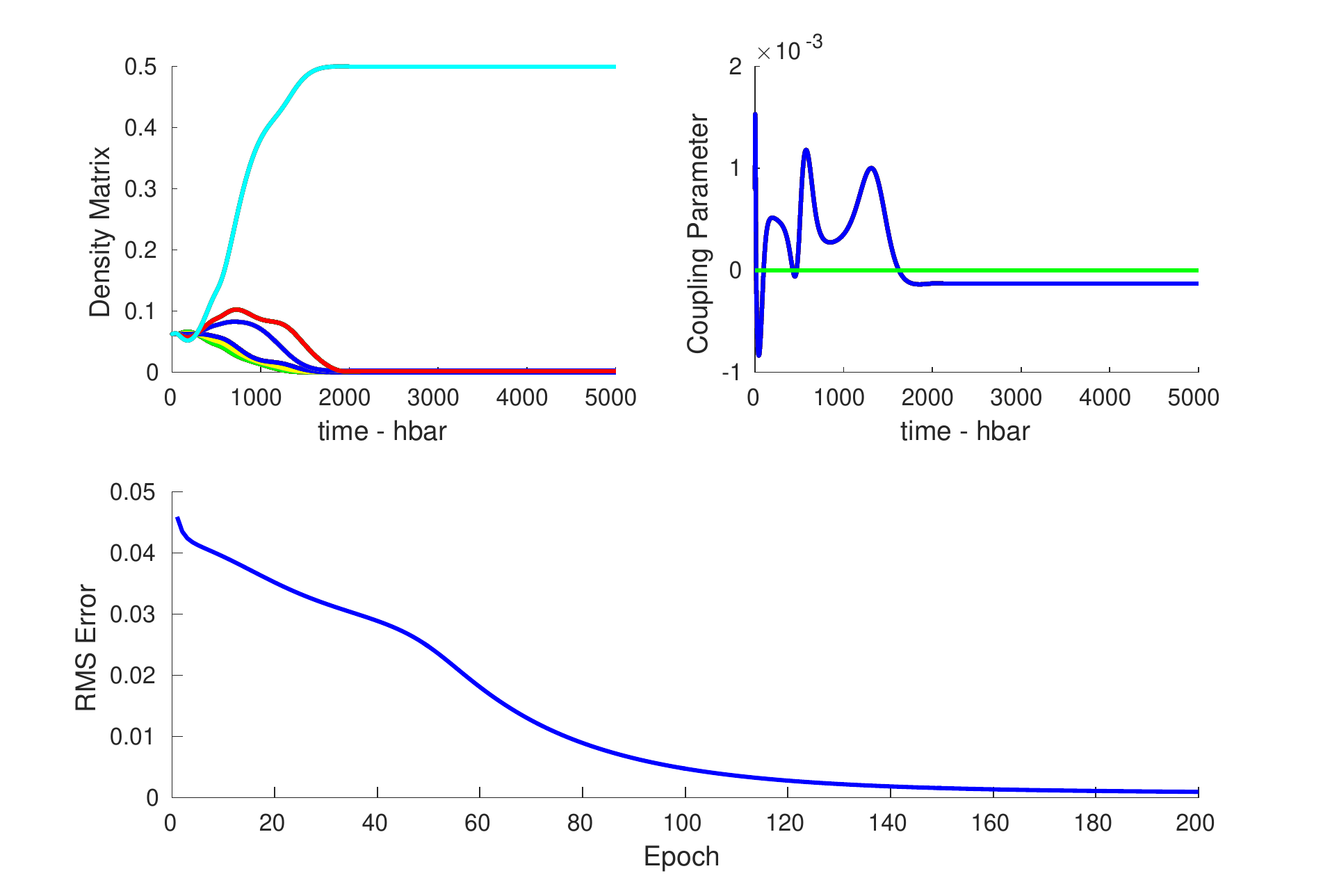, height=2.7in}} %100 percent
\vspace*{13pt}
\fcaption{Training the annealing of the four-qubit system from the flat state to the (fully entangled) GHZ state, starting from the trained three-qubit parameters. The top left graph shows the time evolution of (the absolute magnitude of the elements of) the density matrix $\rho_{I}(t)$, as a function of (annealing) time. Top right shows the trained parameter function $\zeta$, which directs the annealing, as a function of time, in units of GHz. The bottom graph shows the root mean squared error for training, as a function of epoch. The asymptotic error was 0.0009537. The learning rate was $1.25 \times 10^{-05}$.}
\end{figure}

\begin{figure} [htbp]
\centerline{\epsfig{file=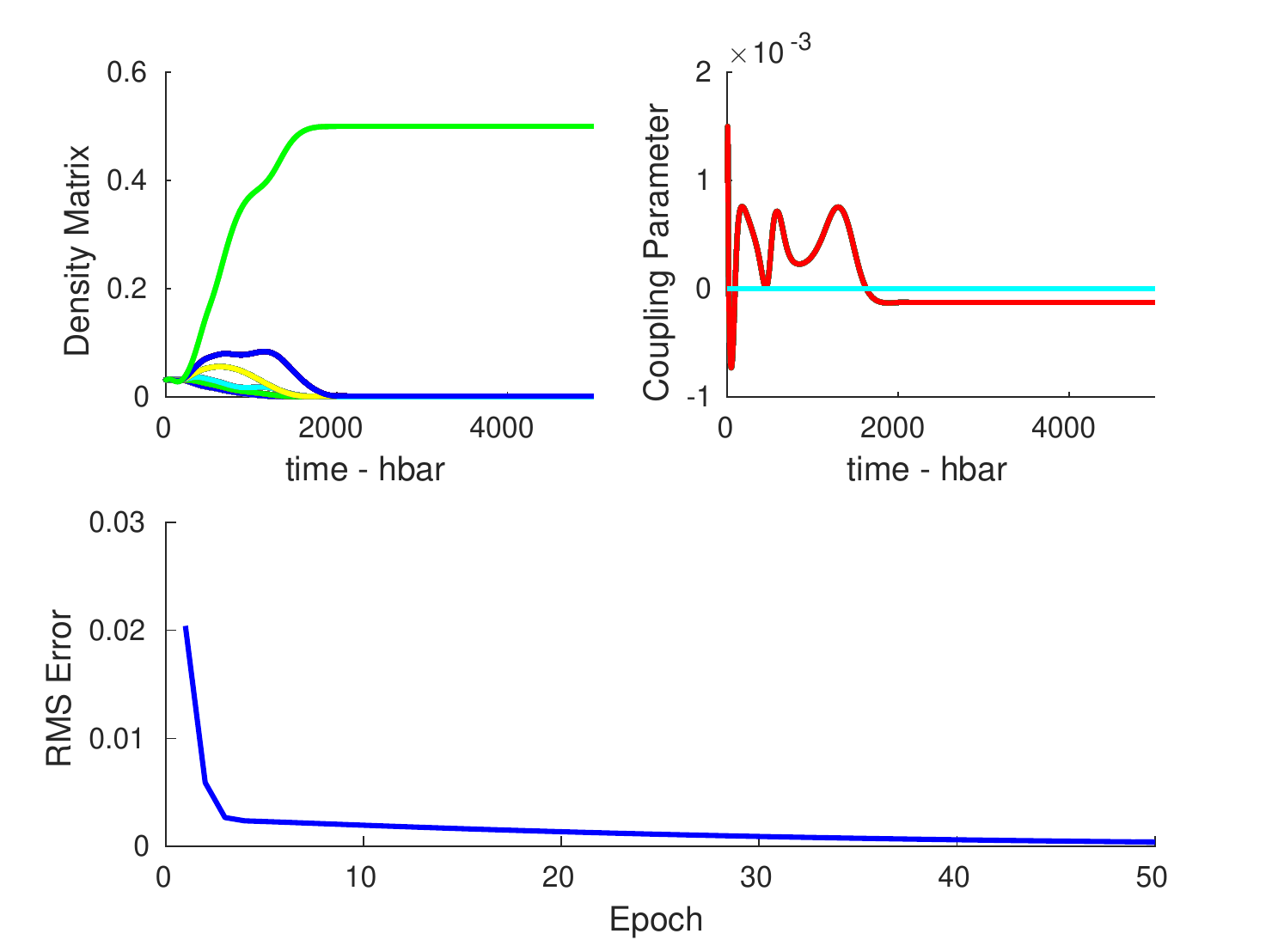, height=2.7in}} %100 percent
\vspace*{13pt}
\fcaption{Training the annealing of the five-qubit system from the flat state to the (fully entangled) GHZ state, starting from the trained four-qubit parameters. The top left graph shows the time evolution of (the absolute magnitude of the elements of) the density matrix $\rho_{I}(t)$, as a function of (annealing) time. Top right shows the trained parameter function $\zeta$, which directs the annealing, as a function of time. The bottom graph shows the root mean squared error for training, as a function of epoch. The asymptotic error was 0.0004119.  The learning rate was $1.25 \times 10^{-05}$. }
\end{figure}

\begin{figure} [htbp]
%\centerline{\epsfig{file=2-3-4-5-6-eps-converted-to.pdf, height=2.7in}} %100 percent
\vspace*{13pt}
\fcaption{Training the annealing of the six-qubit system from the flat state to the (fully entangled) GHZ state, starting from the trained five-qubit parameters. The top left graph shows the time evolution of (the absolute magnitude of the elements of) the density matrix $\rho_{I}(t)$, as a function of (annealing) time. Top right shows the trained parameter function $\zeta$, which directs the annealing, as a function of time, in units of GHz. The bottom graph shows the root mean squared error for training, as a function of epoch. The asymptotic error was 0.002576. The learning rate was $1.25 \times 10^{-05}$. }
\end{figure}

There is quite a bit more variation in $\zeta(t)$ as the system grows in size beyond three qubits. This may be superfluous, though. We found, in previous work on training for an entanglement indicator, that the backpropagation training produced considerable high frequency fluctuation, even though a Fourier series of only one or two terms sufficed to reproduce essentially all the behavior necessary for the indicator\cite{behrmannoise}. Further work on the minimum necessary is ongoing. 

\vspace*{1pt}\textlineskip	%) USE THIS MEASUREMENT WHEN THERE IS
\section{ Broken path from flat to GHZ} %) A SECTION HEADING
\vspace*{-0.5pt}
\noindent

So far we have seen that the system can ``learn'' a coupling parameter function $\zeta(t)$ that would take the system from an initial ``flat'' state $\rho_{\texttt{flat}}= \frac{1}{2^{N}}\prod_{i=1}^{N} [|0\rangle + |1\rangle]_{i}\otimes [\langle 0| + \langle 1|]_{i}$ to a GHZ state $\rho_{\texttt{GHZ}} = \frac{1}{2} [|0..0_{N}\rangle +|1...1_{N}\rangle]\otimes [\langle 0...0_{N}|+\langle 1...1_{N}|]$, using just a time varying coupling function $\zeta$. In simulation it is of course easy to find the error. But in an actual physical experiment, the only measureable quantities are the average values of the spins. If we calculate the average values of the spins during the annealing time, they are always zero. This is not surprising because the intial and final states are both symmetric. Is it possible to choose an annealing path such that the average spins of an intermediate state are not zero, so that there is an experimental check?

We consider first the simple two-qubit case. We choose the evolution from flat to GHZ as taking place along a trajectory which passes through, as intermediate, the state $|Y(\gamma)\rangle = \frac{1}{\sqrt{3+(1-\gamma)^{2}}}[|00\rangle + (1-\gamma)(|01\rangle + |10\rangle + |11\rangle)]$. The tunneling parameter  $K_{A}(t)=K_{B}(t)$ will still not be trained but remain as shown in Figure 1; however, we will now have to train $\varepsilon$ as well as $\zeta$. For the first half of the evolution, we initialize our system in the flat state and train to the partially entangled states $|Y(\gamma)\rangle $, for $\gamma $ between zero (the flat state) and one (the state $|00\rangle$). To speed the training we bootstrap now on the state instead of on the size of the system -- that is,  we start the training process for the parameter functions $\zeta$ and $\varepsilon$, for each value of $\gamma$, from the trained functions for the previous, close value of $\gamma$.  (That is, we bootstrap from $\gamma = 0 \rightarrow 0.1 \rightarrow 0.2 ... \rightarrow 1.0$.)  Figure 7 shows results for the first step of the path: from flat to $|Y(\gamma)\rangle$. The top left figure shows the rms error for each run, as a function of epoch (pass through the training set). As $\gamma$ increases the system needs progressively more training, but even for $\gamma = 1$ (the top curve), training is essentially complete after 50 epochs. The lower left graph shows the trained coupling parameter function $\zeta$; from bottom to top the curves indicate increasing values of $\gamma$. The bottom right graph  shows the trained bias parameters $\varepsilon_{A}$ and $\varepsilon_{B}$ (in blue and green, respectively); again, the curves are for increasing $\gamma$ from bottom to top. The average value of each spin as functions of $\gamma$, which are the measurable quantities at the endpoints of each path, are shown in the top left graph.  

\begin{figure} [htbp]
\centerline{\epsfig{file=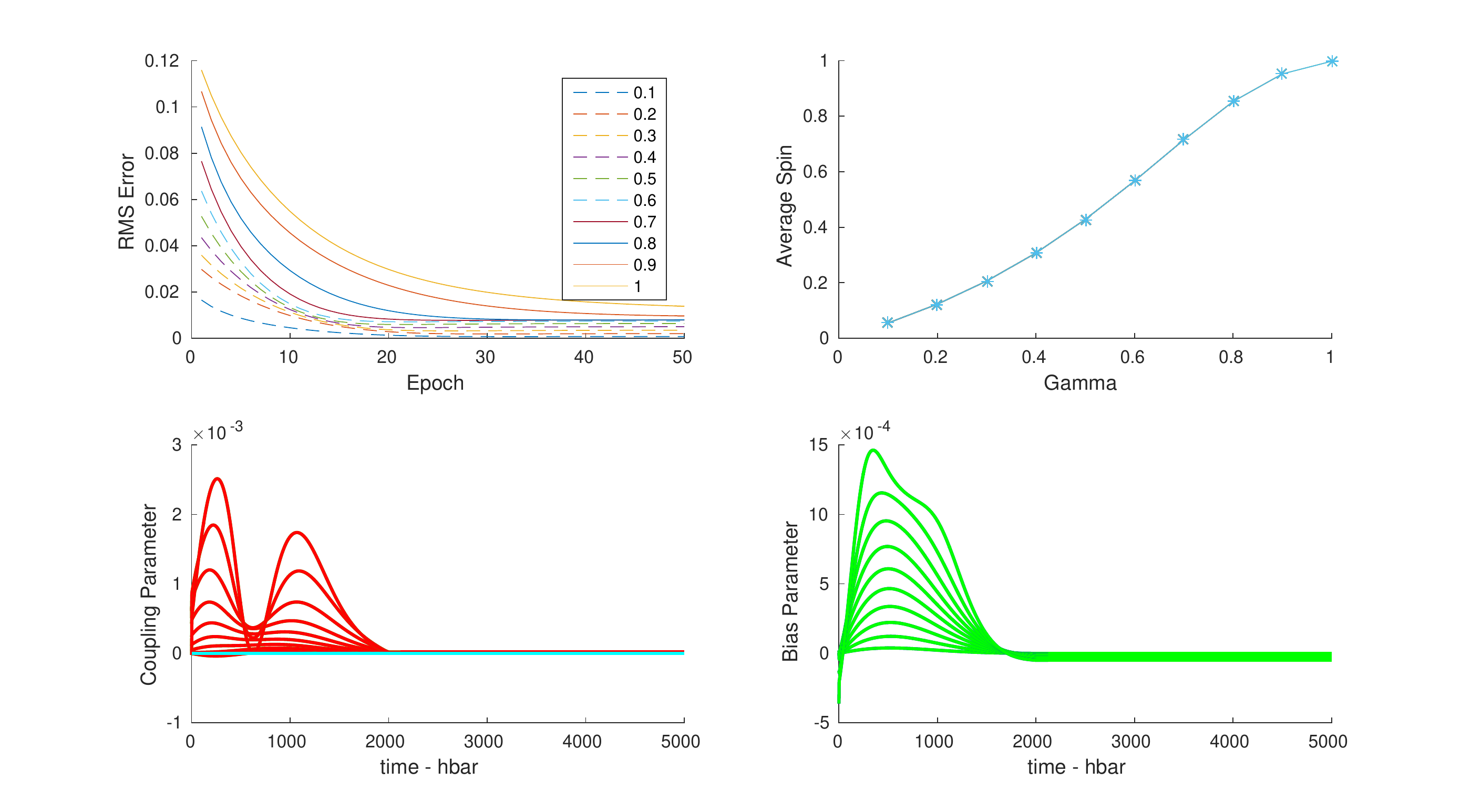, height=2.7in}} %100 percent
\vspace*{13pt}
\fcaption{Training the first step of the broken path from flat to Bell, for the two-qubit system. This first step takes the system from the flat to the state $|Y(\gamma)\rangle = \frac{1}{\sqrt{3+(1-\gamma)^{2}}}[|00\rangle + (1-\gamma)(|01\rangle + |10\rangle + |11\rangle)]$. Counterclockwise from top left, the graphs show: the rms error as a function of epoch, for ten values of $\gamma$; the coupling parameter $\zeta$ for the same values of $\gamma$; the bias parameter $\varepsilon$; and the average value of the spin. Training rates were: $\eta_{\zeta}= 1.25 \times 10^{-05}$, and $\eta_{\epsilon} = 5 \times 10^{-6}$}. 
\end{figure}

In the next figure (Figure 8) we see the corresponding set of graphs for the second part of the time evolution, going from $|Y'(\gamma)\rangle = \frac{1}{\sqrt{1+\gamma^{2}}}[|00\rangle + \gamma|11\rangle)]$ to the Bell state (GHZ for two qubits), $\frac{1}{\sqrt{2}}[|00\rangle + |11\rangle]$. The graphs, counterclockwise from top left, again show the rms error, the trained coupling parameter function $\zeta$; the trained bias paramters $\varepsilon_{A}$ and $\varepsilon_{B}$; and the average value of the spins. Again we make use of bootstrapping to reduce the training time; again, training is essentially complete after only 50 passes through the training set. 

\begin{figure} [htbp]
\centerline{\epsfig{file=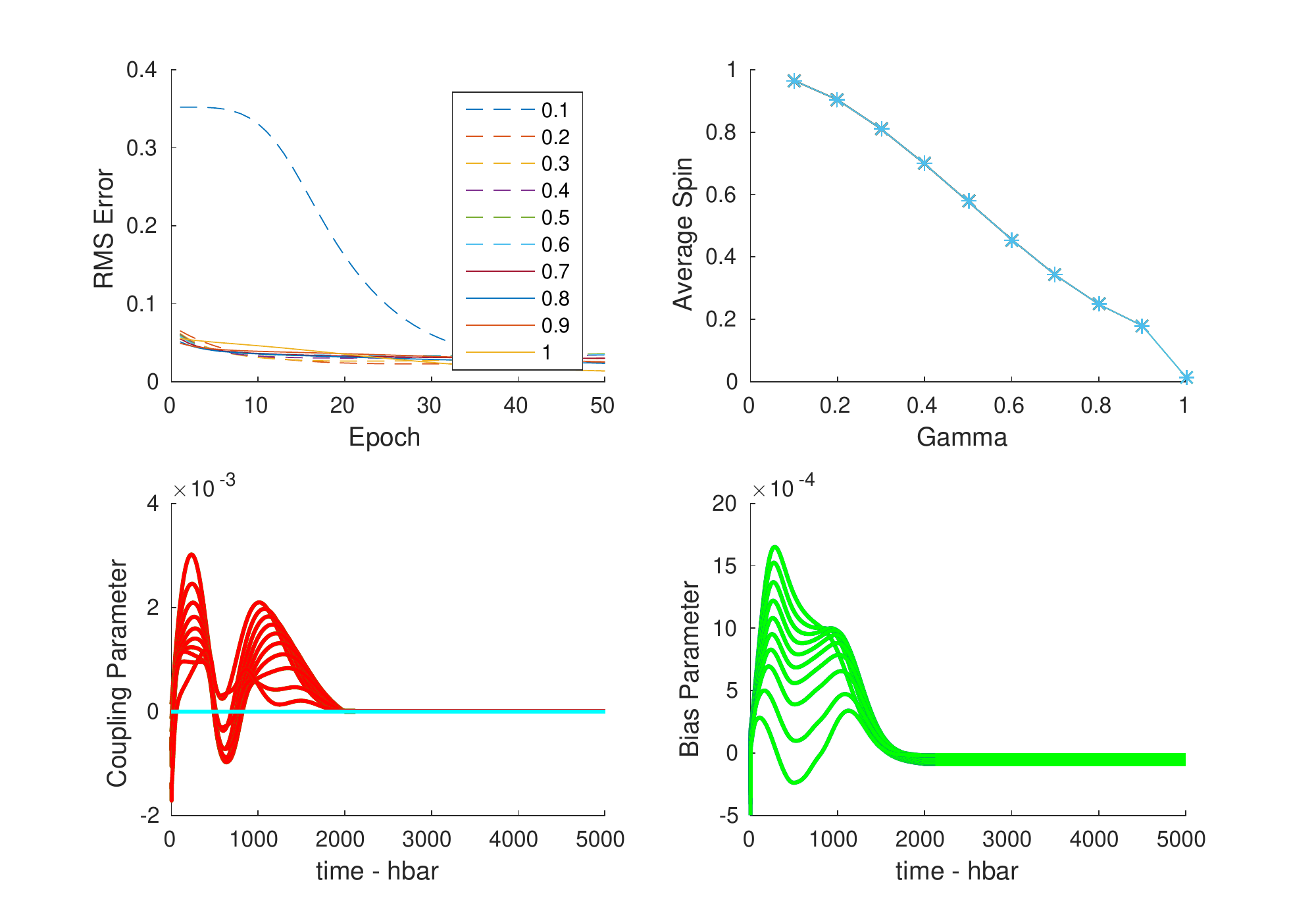, height=2.7in}} %100 percent
\vspace*{13pt}
\fcaption{Training the second step of the broken path from flat to Bell, for the two-qubit system. This second step takes the system from the state $|00\rangle$ to the state  $|Y'(\gamma)\rangle = \frac{1}{\sqrt{1+\gamma^{2}}}[|00\rangle + \gamma|11\rangle)]$. Note that  $|Y'(\gamma=0)\rangle$ is the state $|00\rangle$, and $|Y'(\gamma=1)\rangle$  is the Bell state, $|Bell\rangle = \frac{1}{\sqrt{2}}[|00\rangle + |11\rangle]$. Counterclockwise from top left, the graphs show: the rms error as a function of epoch, for ten values of $\gamma$; the coupling parameter $\zeta$ for the same values of $\gamma$; the bias parameter $\varepsilon$; and the average value of the spin. Training rates were: $\eta_{\zeta}= 1.25 \times 10^{-05}$, and $\eta_{\varepsilon} = 5 \times 10^{-6}$ }.
\end{figure}

The technique can also be extended to larger systems. In Figures 9 and 10 we show results for the three-qubit system, for the broken path from the flat state to the intermediate state $|X(\gamma)\rangle =  \frac{1}{\sqrt{2+6(1-\gamma)^{2}}}[|000\rangle + |011\rangle + (1-\gamma)(|001\rangle + |010\rangle + |100\rangle + |110\rangle + |101\rangle + |111\rangle)]$, and thence to the three-qubit GHZ state, $|GHZ_{3}\rangle = \frac{1}{\sqrt{2}}[|000\rangle + |111\rangle]$. Training for the first step was not as rapid, and required 100 epochs, but the second step was essentially complete by 50 epochs.

\begin{figure} [htbp]
\centerline{\epsfig{file=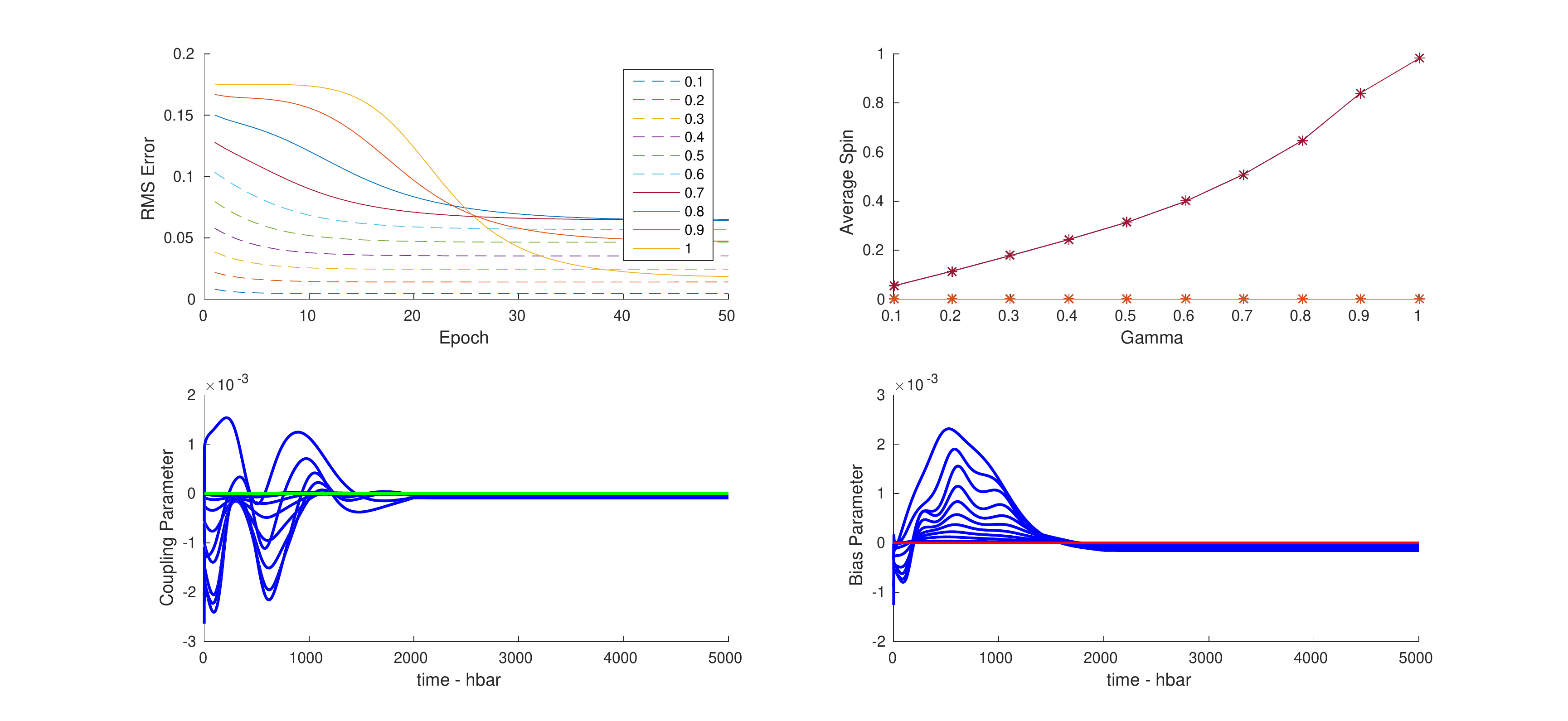, height=2.7in}} %100 percent
\vspace*{13pt}
\fcaption{Training the first step of the broken path from flat to GHZ, for the three-qubit system. This first step takes the system from the flat to the state $|X(\gamma)\rangle =  \frac{1}{\sqrt{2+6(1-\gamma)^{2}}}[|000\rangle + |011\rangle + (1-\gamma)(|001\rangle + |010\rangle + |100\rangle + |110\rangle + |101\rangle + |111\rangle)]$.  Note that $|X(\gamma=0\rangle$ is the flat state, and $|X(\gamma=1)$ is the state $\frac{1}{\sqrt{2}}[|000\rangle + |011\rangle]$. Counterclockwise from top left, the graphs show: the rms error as a function of epoch, for ten values of $\gamma$; the coupling parameter $\zeta$ for the same values of $\gamma$; the bias parameter $\varepsilon$; and the average value of the spin. Training rates were: $\eta_{\zeta}= 1.25 \times 10^{-05}$, and $\eta_{\epsilon} = 5 \times 10^{-6}$}.
\end{figure}

\begin{figure} [htbp]
\centerline{\epsfig{file=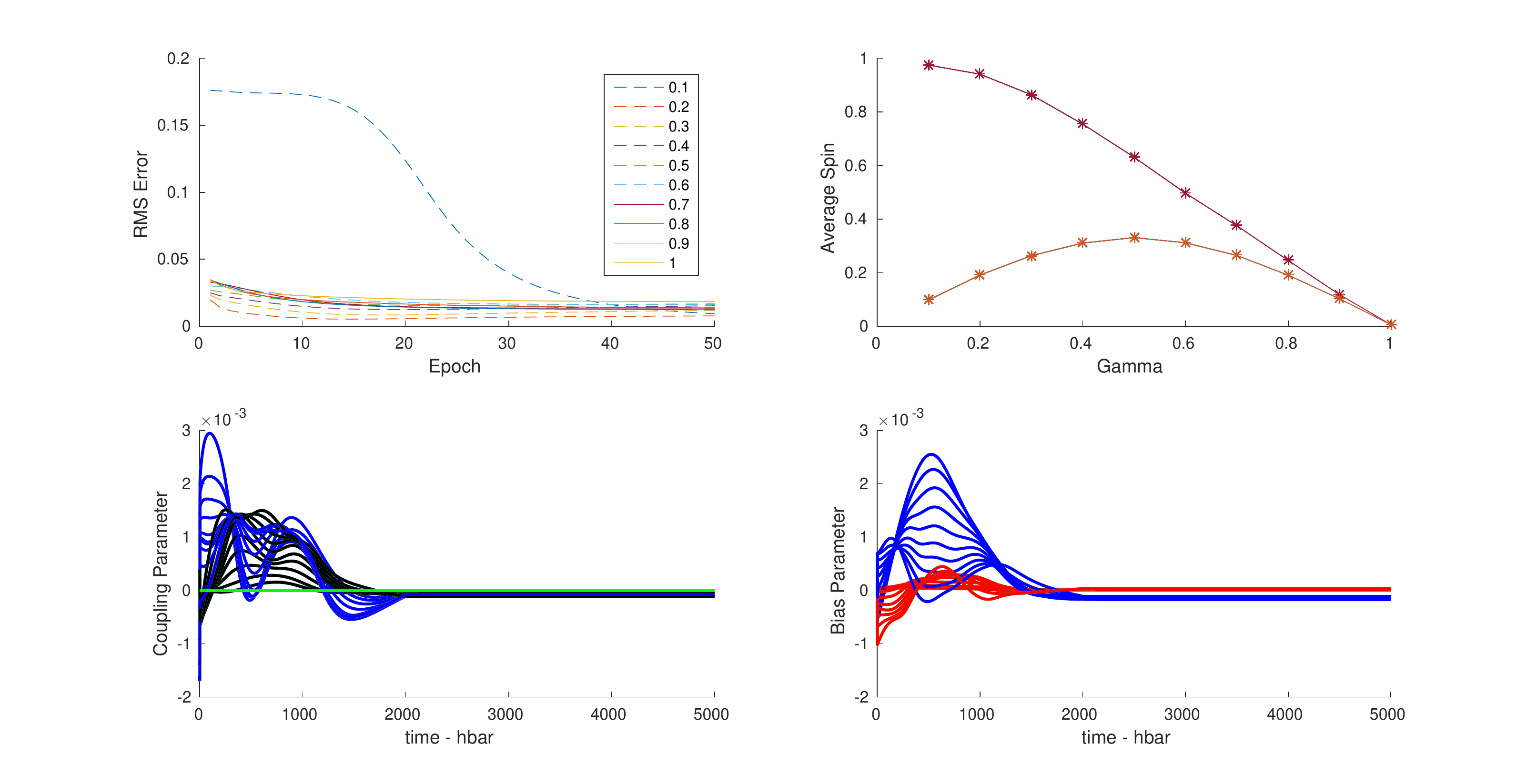, height=2.7in}} %100 percent
\vspace*{13pt}
\fcaption{Training the second step of the broken path from flat to GHZ, for the three-qubit system. This second step takes the system from the state $\frac{1}{\sqrt{2}}[|000\rangle + |011\rangle]$ to the state $|X'(\gamma)\rangle =  \frac{1}{\sqrt{1+ \gamma^{2}+(1-\gamma)^{2}}}[|000\rangle + (1-\gamma)|011\rangle + \gamma|111\rangle)]$.  Note that $|X'(\gamma=1)$ is the GHZ state for the 3-qubit system, $|GHZ_{3}\rangle = \frac{1}{\sqrt{2}}[|000\rangle + |111\rangle]$. Counterclockwise from top left, the graphs show: the rms error as a function of epoch, for ten values of $\gamma$; the coupling parameter $\zeta$ for the same values of $\gamma$; the bias parameter $\varepsilon$; and the average value of the spin. Training rates were: $\eta_{\zeta}= 1.25 \times 10^{-05}$, and $\eta_{\epsilon} = 5 \times 10^{-6}$ }.
\end{figure}
\vspace*{1pt}\textlineskip	%) USE THIS MEASUREMENT WHEN THERE IS
\section{ Learning the W state} %) A SECTION HEADING
\vspace*{-0.5pt}
\noindent
Another possible entangled target state is the W state, in which a single excitation is shared among N qubits. For a 2-qubit system this this the EPR state, $|W_{2}\rangle = \frac{1}{\sqrt{2}}[|01\rangle + |10\rangle]$. Again both the intial (flat) state and the target state are symmetric (average spin zero.) So, we construct our broken pathway. Again we parametrize, this time training to the intermediate state $|V(\gamma)\rangle = \frac{1}{\sqrt{1+3(1-\gamma)^{2}}}[|01\rangle + (1-\gamma)(|00\rangle + |11\rangle +|10\rangle)]$. Results for training for each of the two steps are shown in Figures 11 and 12.   

\begin{figure} [htbp]
\centerline{\epsfig{file=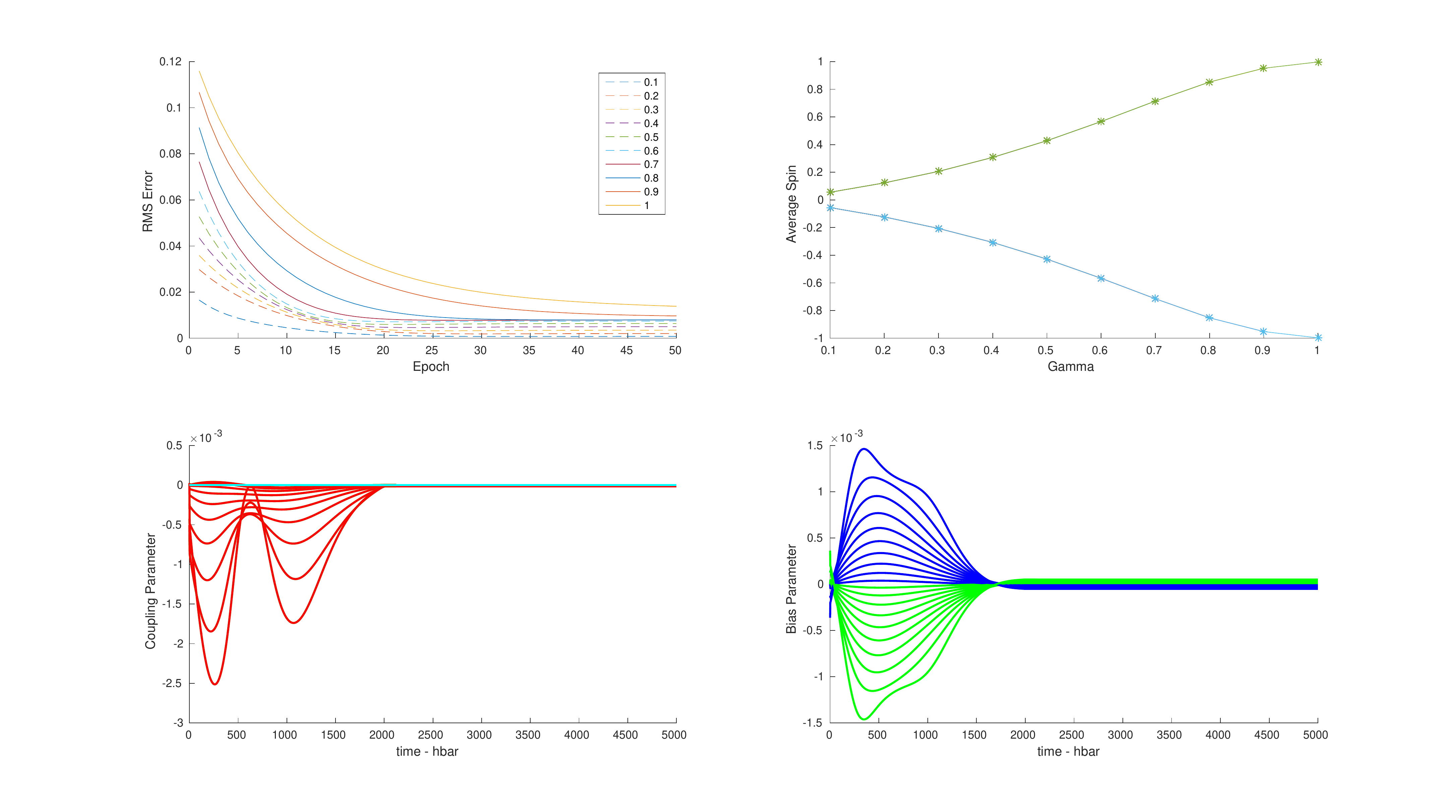, height=2.7in}} %100 percent
\vspace*{13pt}
\fcaption{Training the first step of the broken path from flat to W, for the two-qubit system. This first step takes the system from the flat to the state $|V(\gamma)\rangle = \frac{1}{\sqrt{1+3(1-\gamma)^{2}}}[|01\rangle + (1-\gamma)(|00\rangle + |11\rangle +|10\rangle)]$. Note that $V(\gamma=0)$ is the flat state, and $V(\gamma=1)$ is the state $|01\rangle$. Counterclockwise from top left, the graphs show: the rms error as a function of epoch, for ten values of $\gamma$; the coupling parameter $\zeta$ for the same values of $\gamma$; the bias parameter $\varepsilon$; and the average value of the spin. Training rates were: $\eta_{\zeta}= 1.25 \times 10^{-05}$, and $\eta_{\epsilon} = 5 \times 10^{-6}$}.
\end{figure}

\begin{figure} [htbp]
\centerline{\epsfig{file=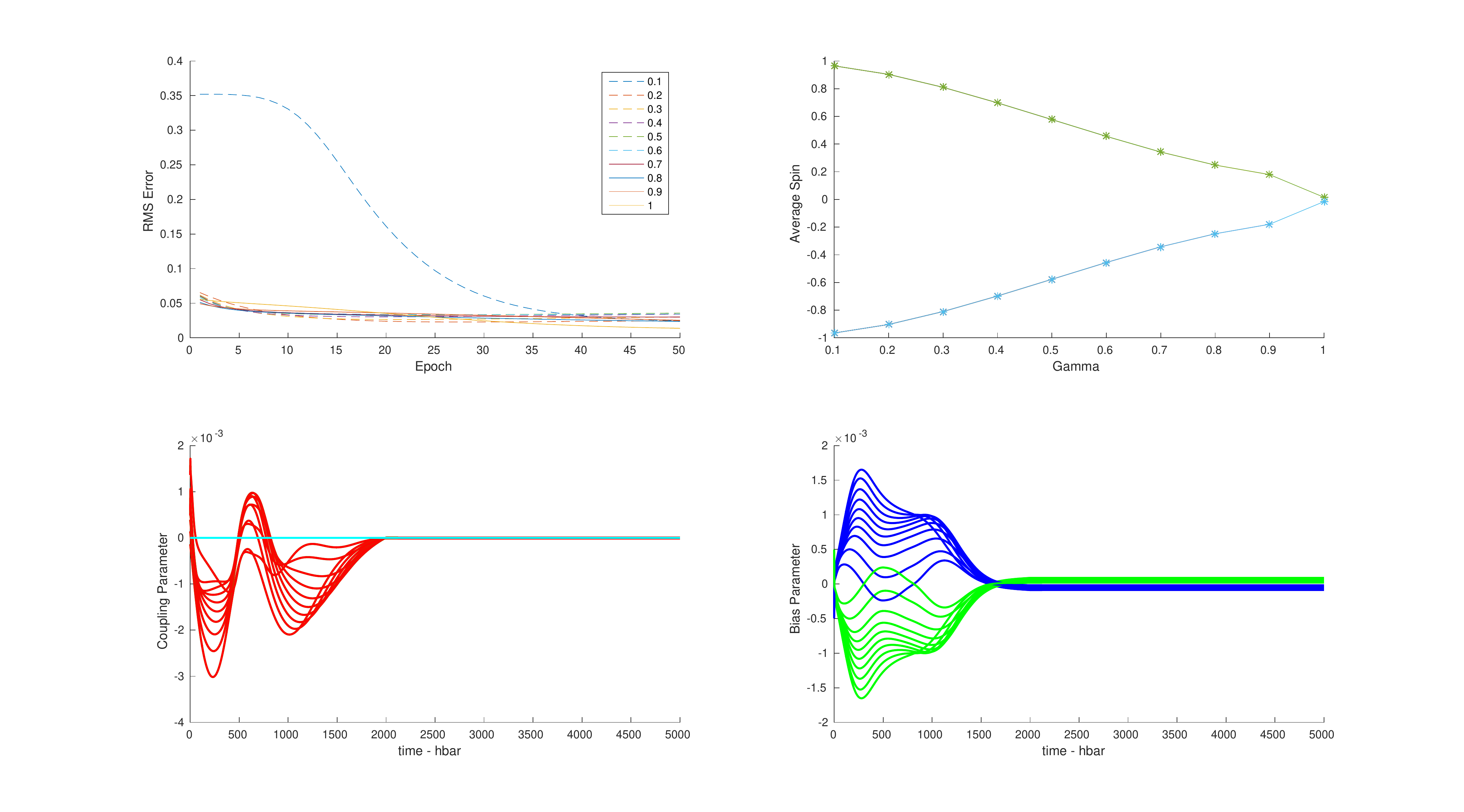, height=2.7in}} %100 percent
\vspace*{13pt}
\fcaption{Training the second step of the broken path from flat to W, for the two-qubit system. This second step takes the system from the state $|V'(\gamma)\rangle = \frac{1}{\sqrt{1+\gamma^{2}}}[\gamma|01\rangle + |01\rangle]$,  to the W state, $|W_{2}\rangle = \frac{1}{\sqrt{2}}[|01\rangle + |10\rangle]$. Note that $|V'(\gamma=0)\rangle$ is the state $|01\rangle$, and $|V'(\gamma=1)\rangle$ is the desired $W_{2}\rangle$ state. Counterclockwise from top left, the graphs show: the rms error as a function of epoch, for ten values of $\gamma$; the coupling parameter $\zeta$ for the same values of $\gamma$; the bias parameter $\varepsilon$; and the average value of the spin. Training rates were: $\eta_{\zeta}= 1.25 \times 10^{-05}$, and $\eta_{\epsilon} = 5 \times 10^{-6}$ }.
\end{figure}

For N greater than two, we need not resort to the broken pathway, since the average final spins are not zero, but to continue illustration of the technique we bootstrap from the two-qubit case, using the analog of the two-qubit $|V_{2}(\gamma)\rangle$ state, $|V_{3}(\gamma)\rangle = \frac{1}{\sqrt{1+7(1-\gamma)^{2}}}[|001\rangle + (1-\gamma)(|000\rangle + |010\rangle +|100\rangle + |011\rangle +|101\rangle + |110\rangle + |111\rangle )]$. Note that $V(\gamma=0)$ is the flat state, and $V(\gamma=1)$ is the state $|001\rangle$. Results for this first step are shown in Figure 13. For the second step training is remarkably easy: a single step gets us to the three-qubit W state. Results are shown in Figure 14. 

For large N, we expect the W states to bootstrap more easily than the GHZ states, because in that limit, the coupling probably only needs to be nearest neighbor. Our earlier work \cite{behrmanmulti} seems to show that for pairwise functions in N qubit systems, there is exponentially less training necessary to go from (N-1) to N qubits, but we have yet to do these calculations on the annealed systems.

\begin{figure} [htbp]
\centerline{\epsfig{file=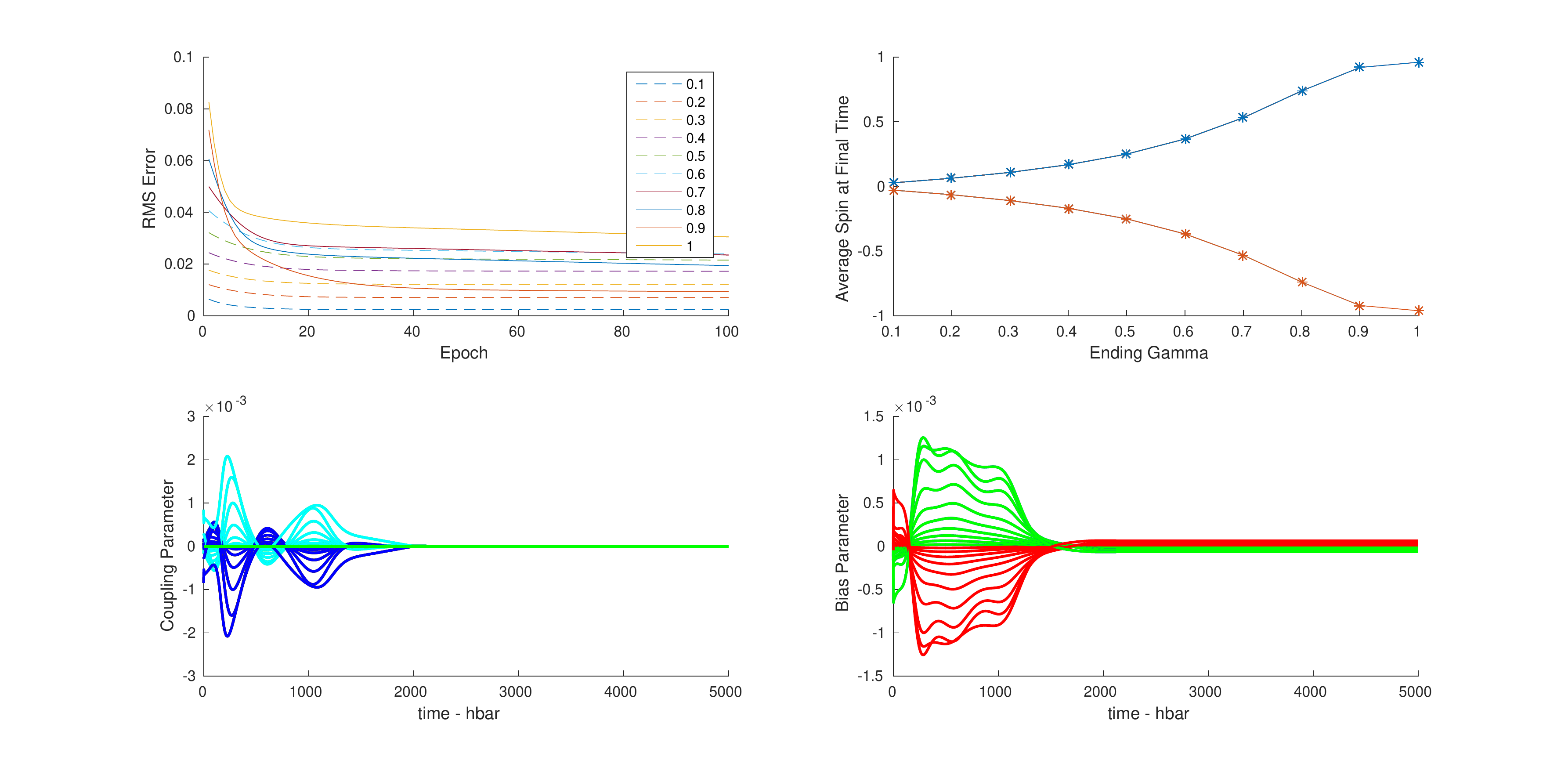, height=2.7in}} %100 percent
\vspace*{13pt}
\fcaption{Training the first step of the broken path from flat to W, for the three-qubit system. This first step takes the system from the flat to the state $|V_{3}(\gamma)\rangle = \frac{1}{\sqrt{1+7(1-\gamma)^{2}}}[|001\rangle + (1-\gamma)(|000\rangle + |010\rangle +|100\rangle + |011\rangle +|101\rangle + |110\rangle + |111\rangle )]$. Note that $V(\gamma=0)$ is the flat state, and $V(\gamma=1)$ is the state $|001\rangle$. Counterclockwise from top left, the graphs show: the rms error as a function of epoch, for ten values of $\gamma$; the coupling parameter $\zeta$ for the same values of $\gamma$; the bias parameter $\varepsilon$; and the average value of the spin. Training rates were: $\eta_{\zeta}= 6.25 \times 10^{-06}$, and $\eta_{\epsilon} = 2.5 \times 10^{-6}$}.
\end{figure}

\begin{figure} [htbp]
\centerline{\epsfig{file=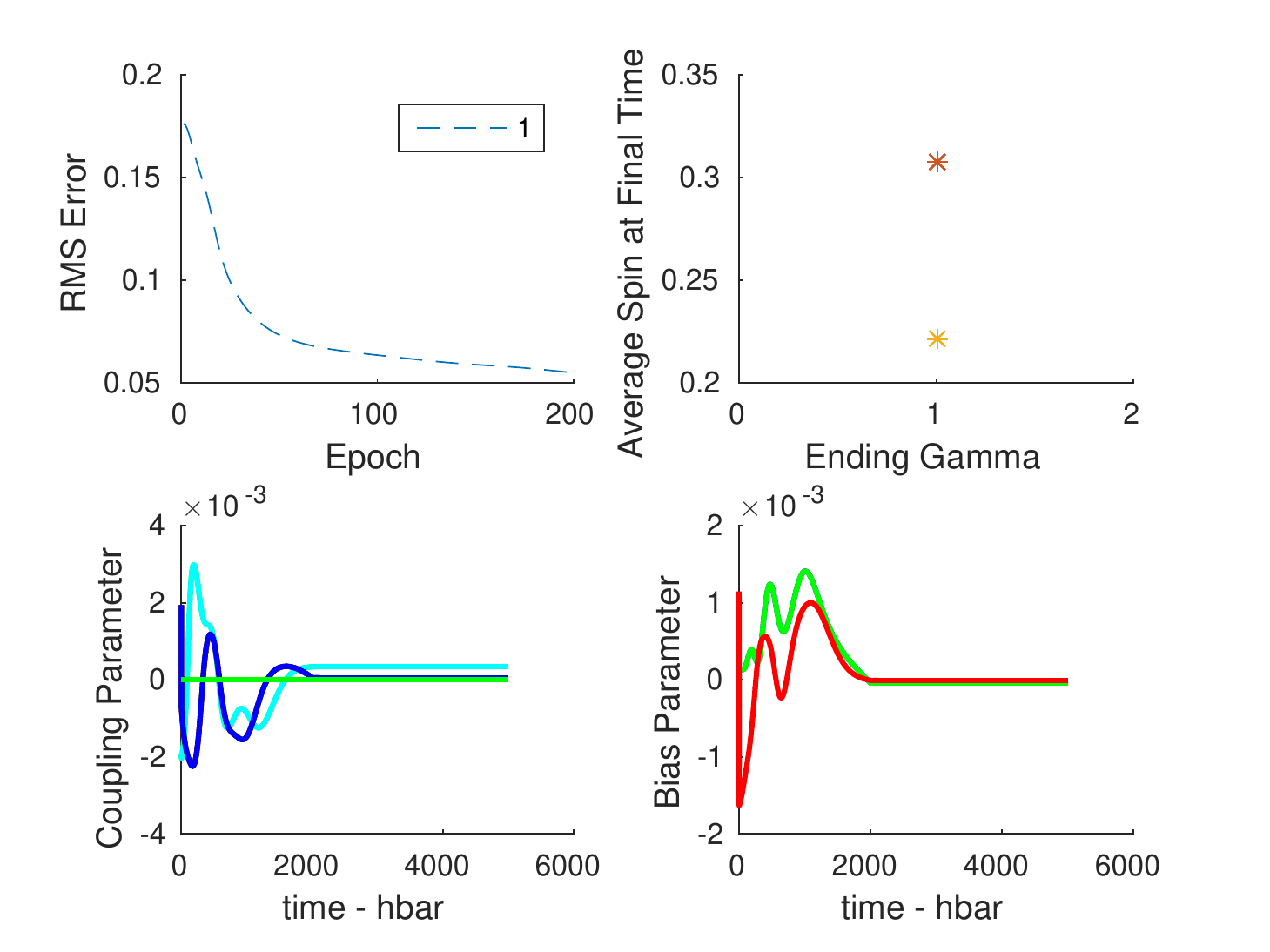, height=2.7in}} %100 percent
\vspace*{13pt}
\fcaption{Training the second step of the broken path from flat to W, for the three-qubit system. This second step takes the system from the state $|001\rangle$  to the W state, $|W_{3}\rangle = \frac{1}{\sqrt{3}}[|001\rangle + |010\rangle + |100\rangle]$. Counterclockwise from top left, the graphs show: the rms error as a function of epoch; the coupling parameter $\zeta$; the bias parameter $\varepsilon$; and the average value of the spin. Training rates were: $\eta_{\zeta}= 6.25 \times 10^{-06}$, and $\eta_{\epsilon} = 2.5 \times 10^{-6}$ }.
\end{figure}

\vspace*{1pt}\textlineskip	%) USE THIS MEASUREMENT WHEN THERE IS
\section{ Robustness to noise and decoherence} %) A SECTION HEADING
\vspace*{-0.5pt}
\noindent

One of the major advantages of a neural network approach is the well-known robustness to incomplete or damaged data. In previous work \cite{behrmannoise} we have shown that this is also true of entanglement indicators as computed by quantum neural networks, and that the QNN method is robust to decoherence, as well. It is only natural to ask, what if the starting state is not exactly the flat state, but contains some small amount of noise as a superposition or admixture of other states? As a first test, on the two-qubit system only, we supposed only that the prepared state has small amounts of (complex) noise in its initial density matrix. Figure 15 shows the root-mean-squared error of a thousand flat states with random noise of differing magnitudes added to the elements of the (initially prepared) two-qubit density matrix, then evolved under the trained parameter functions to the annealed, final state. Unsurprisingly the maximum rms error is approximately linear in noise magnitude, but, interestingly, the slope is less than one, which means that the QNN is robust to small amounts of both decoherence and noise. So, for example, if five percent total error in the Bell state can be tolerated, the size of the total errors in the initial state can be no larger than about five percent; however, if ten percent total final error can be tolerated, as much as eighteen percent noise can be allowed. 

\begin{figure} [htbp]
\centerline{\epsfig{file=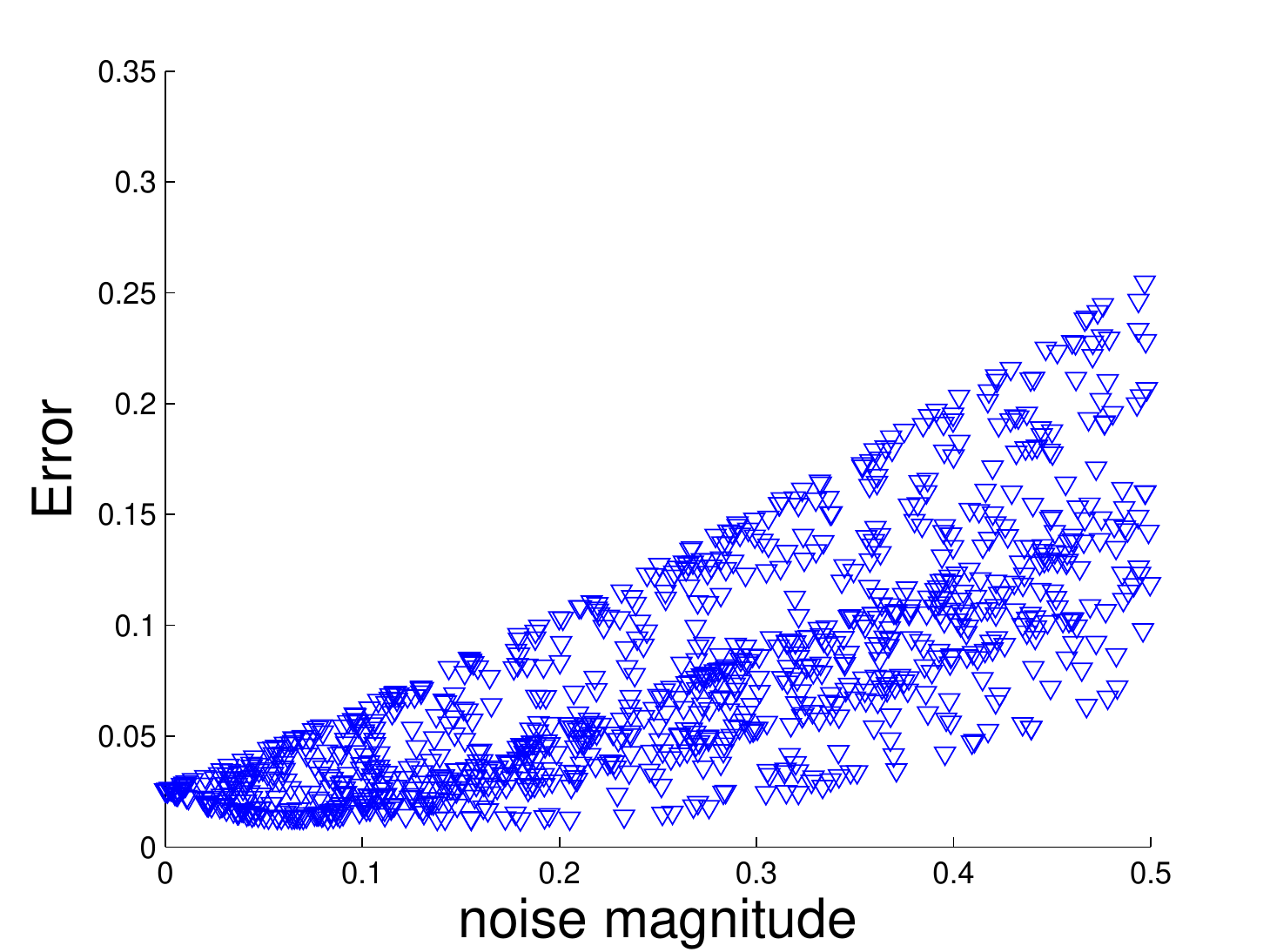, width=8.2cm}} %100 percent
\vspace*{13pt}
\fcaption{Root--mean--squared error for 1000 randomly generated states of the two-qubit system, annealed from the flat state with random complex noise to the Bell state, as a function of total noise magnitude. \label{errorcxnoise}}
\end{figure}

\vspace*{1pt}\textlineskip	%) USE THIS MEASUREMENT WHEN THERE IS
\section{ Complexity and computational cost} %) A SECTION HEADING
\vspace*{-0.5pt}
\noindent

In a series of papers\cite{behrmanqic, behrmanieee,behrmannabic,behrmanmulti,robust} we have explored the possibility of using one or more physical measurements on an N-qubit quantum system as outputs for the QNN; specifically, measurements, at the final time, that would estimate the entanglement of the initial (input) state. For a pure product N-qubit input state (the minimum flexibility in QNN training) it is easy to show that each output of this type can be written as a sum of quadratics in the amplitudes of the input state $|\psi(0)\rangle$, with linearly independent coefficients, plus sums and products of cosines and sines in each of the phase angles in additional cross terms of the amplitudes. Because each of the parameter functions can be taken to be time varying (as we do here), this essentially means that any single measurement output can have the complexity of almost any reasonably well-behaved function, a kind of ``quadratic spline'' in the amplitudes of the elements of the density matrix, and enough nonlinear cross terms to approximate their phase dependence.  And, indeed, we were quite successful in showing that this mapping, for the entanglement indicator, is relatively easy\cite{behrmanqic}, that it bootstraps well to larger systems with a difficulty that decreases with size\cite{behrmanmulti}, and that it is robust to both noise and decoherence, with increasing robustness as the system gets larger\cite{robust}. 

Here, our calculations are a bit different. First, the temporal path of the quantum system is now in complex time, as the temperature and tunneling amplitudes are lowered. We expect that this will increase the stability of the calculations, for the obvious reason that fluctuations will be exponentially suppressed. Second, our current formulation is based on the output's being the quantum state itself. Now, without heroic measures currently not possible on commercially available systems, one cannot determine the quantum state; fortunately, only our learning algorithm needs to know the differences. In the neural network literature, this is called ``offline'' training. We anneal the physical system to a series of intermediate states along our broken pathway, and, for each, perform a measurement of the average spin of each qubit; these measurements are easily possible and provide the needed verification. In terms of the complexity: each of the measurements of an average spin at a qubit site, at each intermediate step along the broken pathway, is an output, and, thus, can have at least the complexity of a sum of N quadratics of the amplitudes plus nonlinear cross terms in the sines and cosines of the phase angles. 

Detailed exact simulations of quantum systems in real time are, of course, computationally expensive. While we are hopeful that this approach will eventually be useful for dealing with systems of an interesting size, we do not, realistically, expect that with just the techniques explored in this, first, paper, we will be able to do exact simulations on thousand-qubit systems.  Several avenues present themselves. We are currently working to reformulate our method so as to enable ``online'' training. With online training the physical system could be initialized in the flat state, then annealed to some intermediate state, using the offline calculations as a guide. The physical measurement would then be performed, errors calculated, and on that basis, the annealing parameter functions for that step would be modified. The cycle would be repeated until the measured errors were as small as desired. Once that was accomplished, annealing could take place from the flat to the second step, and so on.

\vspace*{1pt}\textlineskip	%) USE THIS MEASUREMENT WHEN THERE IS
\section{ Conclusions} %) A SECTION HEADING
\vspace*{-0.5pt}
\noindent
We have shown that for a multi qubit system it is possible systematically to find parameters such that the system anneals reproduceably to a fully entangled state, a kind of quantum control. We have also shown that it is possible to ``bootstrap'' in two ways: first, from a smaller system to a larger one, and, second, from one final desired state to another that is close by, using the knowledge of the previously learned parameter functions. In general the amount of additional training diminishes with increasing size, which raises hopes for the applicability of our technique to systems even of hundreds of qubits. The exception was in the training from three qubits to four (possibly for symmetry reasons - we saw this phenomenon also with the entanglement indicator\cite{behrmanmulti}), though even here, the initial error for the four-qubit calculation was a tenth the size when starting from the three-qubit functions, compared with starting from scratch. And, while direct training by symmetry occurs along a pathway whose progress is not experimentally accessible, we have shown that it is possible to specify a pathway along which the average spin is nonzero and thus checkable.

We are also working to tailor our annealing process specifically to a commercially available setup, for which the current interface does not allow the parameters to be varied independently. We define a monotonically increasing annealing parameter, $S{w}$, which is what is trained; the parameters of the Hamiltonian, $\varepsilon$, $\zeta$, and $K$ depend on $S_{w}$: 
\noindent
\begin{equation}
\zeta(t) = \frac{S_{w}(t) - S_{w}(0)}{S_{w}(t_{kf})-S_{w}(0)} \zeta(t_{kf})
\end{equation}
and similarly for  $\varepsilon$ and $K$. Annealing occurs during the time from t=0 to
t=$t_{kf}$. The initial values of the parameters are fixed at t=0, and the final values at the end of annealing are adjusted via training as well, except for the tunneling parameter $K$, which must be zero at the end.  Additionally, $S_{w}$ is forced to be monotonically increasing. Training of the 2-qubit system for 1000 epochs results in time histories of $\zeta$   that force the system to anneal to the target entangled density matrix, as we saw before. Figure 16 shows the training of the annealing from the flat state to the Bell state, analogous to Figures 1 and 2. Considerably more training was necessary with the restrictions imposed, and bootstrapping is also more difficult; however, just as with our previous work on entanglement indicators\cite{robust}, it seems that much of the variation that straightforward training produces in the coupling is unnecessary. We are currently working on more sophisticated methods of optimizing the training times and rates.

\begin{figure} [htbp]
\centerline{\epsfig{file=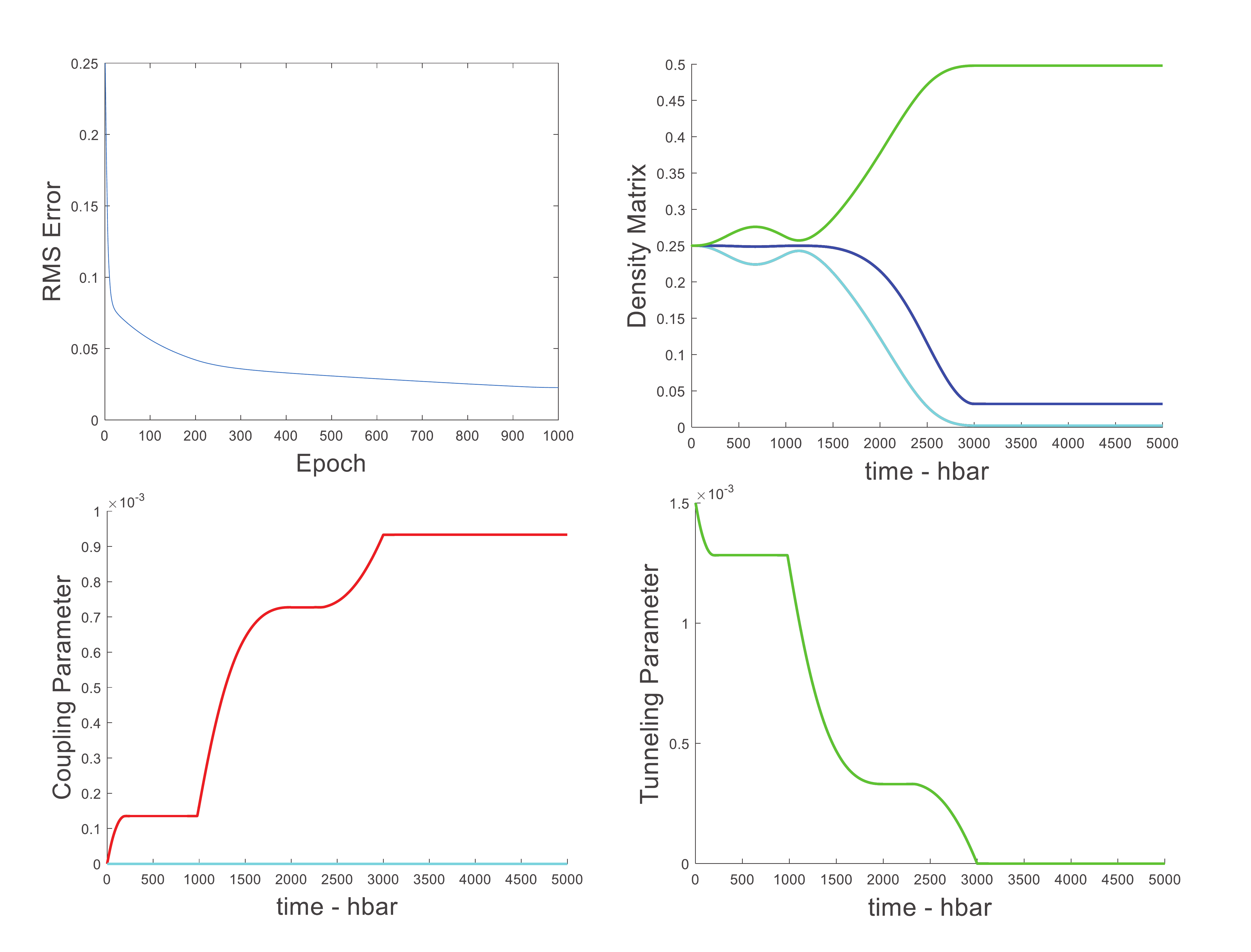, height=2.7in}} %100 percent
\vspace*{13pt}
\fcaption{Training the 2-qubit system from flat to Bell, using a single monotonic function $S{w}$.  Counterclockwise from top left, the graphs show: the rms error as a function of epoch; the coupling parameter $\zeta$; the tunneling parameter, $K$; and the density matrix during annealing. The bias parameter $\varepsilon$ was zero for all times and epochs. Training rates were: $\eta_{\zeta}= 1.25 \times 10^{-05}$, and $\eta_{\epsilon} = 5 \times 10^{-6}$}. 
\end{figure}

\nonumsection{Acknowledgements}
\noindent
We are grateful for illuminating discussions with Adrian Keister and Nam Nguyen of WSU, and with Trevor Lanting and Murray Thom of DWave.

\end{document}